\documentclass[options]{JHEP3}

\usepackage{epsfig}

\def\ra{{\rightarrow}}
\def\a{{\alpha}}
\def\b{{\beta}}

\def\eps{{\epsilon}}

\def\pr{{\partial}}

\def\sp{\vspace{.15in}}

\newcommand{\be}{\begin{equation}} \newcommand{\ee}{\end{equation}}
\newcommand{\bea}{\begin{eqnarray}}\newcommand{\eea}
{\end{eqnarray}}

\title{Noncommutative brane-world, (Anti) de Sitter vacua\hspace{.3in}{} and extra dimensions}
\author{{Supriya Kar}\\ Department of Physics and Astrophysics, University of Delhi, Delhi 110 007, India\\ E-mail: \email{skkar@physics.du.ac.in}}

\abstract{We investigate a curved brane-world, inspired by a noncommutative $D_3$-brane, in a type IIB string theory.
We obtain, an axially symmetric and a spherically symmetric, (anti) de Sitter black holes in $4D$. 
The event horizons of these black holes possess a constant curvature and may be seen to be governed by different 
topologies. The extremal geometries are explored, using the noncommutative scaling in the theory, to reassure the 
attractor behavior at the black hole event horizon. The emerging two dimensional, semi-classical, black hole is 
analyzed to provide evidence for the extra dimensions in a curved brane-world. It is argued that the gauge nonlinearity 
in the theory may be redefined by a potential in a moduli space. As a result, $D=11$ and 
$D=12$ dimensional geometries may be obtained at the stable extrema of the potential.}

\preprint

\begin{document}

\section{Preliminaries}

\subsection{Introduction}

In the recent years, considerable amount of interest has been devoted to the construction of 
meta stable de Sitter (dS) vacua in quantum gravity \cite{witten1}-\cite{strominger2}. The primary difficulty
is due to the fact that the complete event horizon in dS black hole is not accessible to an observer
due to its hyperbolic geometry. Nevertheless, the construction 
of dS vacua has been achieved by taking into account a small number of $D_3$-branes along with the AdS vacua 
in a type IIB string theory \cite{kklt}. On the other hand, 
anti de Sitter (AdS) spaces are well understood in a string theory due its established correspondence with the
gauge theory at its boundary \cite{maldacena,witten2}.

\par
\sp
In the context, the the nonlinear electromagnetic (EM-) field on a $D_3$-brane turns out
to be a potential candidate to address some of the quantum aspects of gravity \cite{gibbons-ishi}.
In fact, a consistent noncommutative deformations of Einstein gravity has been the subject of interest in the 
recent literature \cite{jevicki}-\cite{kar06}. For a recent review see \cite{sazbo}.

\par
\sp
Very recently, we have obtained an extremal dS geometry by analyzing the tunneling between AdS and 
dS black holes in a curved $D_3$-brane
frame-work \cite{kar06}. In this paper,  we incorporate the formalism of a noncommutative brane-world 
to construct various dS and AdS black holes with different topologies of its event horizons.
In particular, we revisit a curved brane-world \cite{km-2} underlying a noncommutative $U(1)$ gauge theory 
on a $D_3$-brane \cite{seiberg-witten}. We obtain generalized $dS_4$ and $AdS_4$ geometries leading to  
an axially symmetric and a spherically symmetric black holes. The solutions are shown to be characterized by
the effective parameters such as ADM mass and electric or magnetic charge, in addition to the cosmological
constant $\Lambda$ and topology determining parameter $k$. The effective mass and charge 
are argued to incorporate all order corrections in the noncommutative parameter $\Theta$. 
We analyze the gravity decoupling regime 
initiated by the Hawking radiation phenomenon from the generalized black holes. A noncommutative scaling 
\cite{km-1} limit, generated in the frame-work, is explored to obtain the low energy gravity regime. In fact,
the decoupling regime is analyzed carefully to describe various extremal monopole black hole geometries in 
presence of a cosmological constant. It is shown that the extremal geometries lead to govern a $2D$ large 
dS black hole in the frame-work, as the multiple gauge charges reduce the radius of $S^2$ to a vanishingly
small value. A relation between the scalar moduli and the EM-charges in the theory is established at the 
event horizon of the black holes to reconfirm the attractor behavior there. The extremal black holes are
analyzed for the effective gravitational potential to argue for the presence of extra dimensions in a curved
brane-world. The Hawking radiation phenomenon, leading to a series of geometric transitions, is analyzed to
argue for the existence of $D=11$ and $D=12$ extremal geometries. They may correspond to various relevant 
extrema of an effective potential governing the gauge nonlinearity in the moduli space. 

\par
\sp
We plan the paper as follows. In section 1.1, we present the preliminaries to noncommutative brane-world
formalism. The generalized axially symmetric and spherically symmetric static $4D$ black holes are obtained 
respectively in section 2.1 and 2.2. Hawking radiation leading to extremal geometries are discussed
in section 3.1 and 3.2. The hint for large extra dimensions leading to a $5D$ space-time is conceived 
in section 3.3. and for higher dimensions is described in section 3.4. Finally, we conclude the paper in
section 4.

\subsection{Noncommutative brane-world}

We begin by considering a $D_3$-brane in the background of an $5D$ generalized black hole geometries
\cite{km-2}. Following the prescription, we consider the brane as the boundary of the $AdS_5$ geometry. 
In fact, the interpretation has been shown to be in precise agreement with the AdS/ noncommutative gauge 
theory correspondence \cite{li-wu,hashi-itzhaki,ho-li}. The
induced metric $g_{\mu\nu}$ on the boundary $\pr {\cal M}$ is not completely constant, rather it takes into account the dynamical aspects of gravity in the bulk ${\cal M}$. 
In other words, the boundary dynamics is not independent
of that in the string bulk. However, for simplicity we consider a constant $g_{\mu\nu}$ and a constant induced antisymmetric field $b_{\mu\nu}$ on the brane. Then, the $D_3$-brane dynamics can be approximated by
\bea
&&S_{D_3}= - T_{D_3}\int_{\pr{\cal M}} d^4x\ \sqrt{ g+ \bar{\cal F} }\nonumber\\
&&\quad\quad = - \int_{\pr{\cal M}} d^4x\ \sqrt{g} \left (\ T_{D_3}  - {1\over4} {\cal F}^2 \ +\ 
{1\over8} {\cal F}\ {\cal F}_+\ {\cal F}_-^2\ K^2({\cal F})\ \right )\ ,\label{db1}
\eea
where $T_{D_3}$ is the $D_3$-brane tension and $g \equiv \det g_{\mu\nu}$.
The $U(1)$ gauge invariant field strength can be expressed as 
${\cal F}_{\mu\nu} = (2\pi\a')^{-1} b_{\mu\nu} + F_{\mu\nu}$. In addition,
${\cal F}_{\pm}=({\cal F}\pm {}^*{\cal F})$ and $K({\cal F})$ takes account of all the 
higher order ${\cal F}$-terms in the action. The Minkowski inequality in the theory incorporates
the self-duality condition between the EM-fields, $i.e.\ |E| = |B|$, in the world-volume theory.
As a result, the brane dynamics turns out to be exact to all orders in ${\cal F}$. Then, the brane action
can be re-expressed as
\be
S_{D_3}= - \int_{\pr{\cal M}} d^4x\ \sqrt{g} \left (\ \lambda_b  - {1\over4} F_{\mu\nu} F^{\mu\nu} \ \right )
\ ,\label{db2}
\ee
where $\lambda_b$ is a constant.
In fact, $\lambda_b$ can be identified with the brane tension $T_{D_3}$ 
in absence of the $b$-field. However in presence of a significant $b_{\mu\nu}$, the second term in the action (\ref{db1}) can be seen to dominate over the first there, In that case, the constant may be expressed as
$\lambda_b= -{1\over4} b_{\mu\nu}b^{\mu\nu}$.
Often $\lambda_b$ is ignored in the gauge theory, as it does not lead to any dynamics. However, 
in presence of the dynamical gravity, $i.e.$ in a curved $D_3$-brane frame-work, it can be seen 
to be associated with a significant notion of a (bare) cosmological constant.

\sp
\noindent
Interestingly, the analysis for the $U(1)$ gauge dynamics (\ref{db2}) goes through a noncommutative $D_3$-brane.
Using the Seiberg-Witten map \cite{seiberg-witten}, the boundary dynamics can be transformed to
a noncommutative $U(1)$ gauge dynamics on a $D_3$-brane. Then, the noncommutative gauge theory on the $D_3$-brane can be approximated by the Dirac-Born-Infeld dynamics.
It is given by
\be
S_{D_3}= -\int_{\pr{\cal M}} d^4x\ \sqrt{G} \left (\ \lambda_b  - {1\over4} G^{\mu\lambda}
G^{\nu\rho}\ {\hat F}_{\mu\nu}\star {\hat F}_{\lambda\rho} \ \right )\ ,\label{db3}
\ee
where $\lambda_b$ is the brane tension and $G \equiv \det G_{\mu\nu}$. The Moyal $\star$-product accounts 
for the non-locality, which is due to the infinite number of derivatives there. Importantly, the gravitational 
back reaction has been incorporated into the effective theory, which is apparent from the definition of 
the modified metric 
\be
G_{\mu\nu} = (g_{\mu\nu} - [ b g^{-1} b ]_{\mu\nu} + [ b g^{-1}b\ bg^{-1} b]_{\mu\nu} 
+ \dots )\ .\label{dbn3}
\ee
Using the modified metric, the noncommutative gauge potential and 
the corresponding EM-field can be seen to receive higher order corrections in $b$-field.
Explicitly the $B$ or $E$-field in the effective theory is given by
\be
{\tilde E} = \left [{{ E^2}\over{1 + E^2}}\right ]^{1/2}\ .\label{dbn4}
\ee
Since a $D$-brane governs the boundary dynamics of an open string,
the relevant gravity dynamics in its bulk may be incorporated into a curved brane
along with the gauge dynamics of a $D_3$-brane. 
The formulation inspires one to seek for a fundamental theory \cite{vafa}
in presence of a three brane, such as D=12 constructions \cite{tseytlin}-\cite{hewson}.
However our starting point, in this paper, 
lies in the bosonic sector of  $D=10$ type IIB string theory on $K_3\times T^2$. 
Ignoring the Chern-Simons terms, the relevant $4D$ effective string dynamics
in Einstein frame may be given by
\be
S_{\rm st}= - \int d^4x \ {\sqrt{{\tilde G}^E}} \ \bigg ( {1\over{16\pi G_N}} R- 2(\pr\phi)^2
-{1\over{2}} F_1^{(k)} C_{kl} F_1^{(l)} - {1\over{2\cdot 2!}} F_2^{(i)}D_{ij}F_2^{(j)}
- {{Z}\over{2\cdot 4!}} F_4^{(m)}L_{mn}F_4^{(n)} \bigg )\ ,\label{db01}
\ee
where ($C_{kl}$, $D_{ij}$,$L_{mn}$) govern the appropriate moduli coupling to the gauge field of various ranks
and $Z$ is a normalization constant. Then, the nontrivial four form energy density can be given by
a potential in the moduli space
\be
V_4(\phi) ={{Z}\over{48}}F_4^{(m)}L_{mn}F_4^{(n)}\ .\label{db02}
\ee
Now, the curved $D_3$-brane dynamics is obtained by coupling the noncommutative $D_3$-brane 
(\ref{db3}) to an effective string theory (\ref{db01}).
In a static gauge, the complete dynamics of a curved $D_3$-brane can be given by
\be
S= - \int d^4x\ {\sqrt{G}}\ \left (\ {1\over{16\pi G_N}}(R - 2\Lambda ) -2(\pr\phi)^2 -
{1\over2} F_1^{(k)}C_{kl}F_1^{(l)}
-{1\over4} D_{pq}F_2^{(p)}\star F_2^{(q)} \right )\ ,\label{db6}
\ee
where
\be
\Lambda (\phi) = 8\pi G_N \left (\ V_4(\phi) - \lambda_b\ \right )\ .
\qquad\qquad\qquad {}\label{db7}
\ee
$\lambda_b$ can take a large constant value as it can be seen to be controlled by an $U(1)$ gauge 
non-linearity in the theory.
The multiple four forms in the theory together with the brane tension, 
redefine the vacuum energy (\ref{db7}). Since an explicit membrane dynamics is absent in the frame-work
(\ref{db6}), the (multiple) four form equations of motion are worked out to yield
\be
\pr^{\mu}\left ( \sqrt{G}\ L_{mn} F^{(n)}_{\mu\nu\lambda\rho}\right ) = 0\ .\label{db71}
\ee
For stable minima in $V_4(\phi)$, the $L_{mn}$ takes a constant value. 
Then, the solutions to the equations of motion are
given by 
\be
F^{(n)}_{\mu\nu\lambda\rho} = \lambda^{(n)}\eps_{\mu\nu\lambda\rho}\ ,\label{db72}
\ee
where $\lambda^{(n)}$ are constants and 
$\eps_{\mu\nu\lambda\rho}$ is a totally antisymmetric tensor. For stable minimum,
$\Lambda(\phi)$ takes a constant value and is given by
\be
\Lambda(\phi)\rightarrow 8\pi G_N 
\left ( {{Z}\over2} \sum_{n'=1}^{n} \left [\lambda^{(n')}\right ]^2\ -\ \lambda_b\right ) .\label{db8}
\ee
It implies that the multiple four-forms along with the
gauge non-linearity could possibly reduce the effective cosmological constant (\ref{db8}) 
to a small value in $4D$. In fact, a constant $\Lambda$ can also be argued,
when the moduli fields $L_{mn}$ are attracted toward a fixed point, $i.e.$ at the event horizon of
the black hole, in a curved $D_3$-brane frame-work. The equations of motion for the effective gravity, 
scalar and multiple gauge fields are worked out, respectively, to yield
\bea
&&R_{\mu\nu} -\ \Lambda G_{\mu\nu} =\ 8\pi G_N\left ( 4 \pr_{\mu}\phi \pr_{\nu}\phi\ +\
T_{\mu\nu}\ +\ D_{ij}\ T^{(ij)}_{\mu\nu} \ \right )\ ,\nonumber\\
&&\partial_{\mu}\left ( \sqrt{G} \
\partial^{\mu}\phi\right )=\ {1\over{16}}{\sqrt{G}}\ {{\partial
D_{ij}}\over{\partial\phi}}F_2^{(i)}F_2^{(j)} \nonumber\\
{\rm and}&& \partial_{\mu}\left (
{\sqrt{G}}\ D_{pq}\ {\hat F}^{(q)\mu\nu}\right ) = 0\ .\label{db10}
\eea
Where $T_{\mu\nu}$ and $T^{(ij)}_{\mu\nu}$, respectively, denote the energy-momentum tensors
in the $D_3$-brane and in the effective string theories. In principle, 
the noncommutative $D_3$-brane can be replaced by its ordinary counter-part 
to describe a curved $D_3$-brane. A priori, the action (\ref{db6})
with $G_{\mu\nu}\ra g^E_{\mu\nu}$ and ${\hat F}_{\mu\nu}\ra F_{\mu\nu}$ can be
seen to accommodate an additional term, which is given by 
\be
S_g ={1\over4} b_{\mu\nu} \int d^4x\sqrt{g^E} A^{\nu}\pr^{\mu}(g^E)_{\lambda\rho}( g^E)^{\lambda\rho}\ .
\label{db1001}
\ee
However, the equations of motion (\ref{db10}) are not modified with the ordinary $U(1)$ gauge dynamics
in the curved brane theory.
In general ${\hat Q}^{(m)}= (1/4\pi)\int_{S^2} {}^{\star}F_2^{(m)}$ denote the effective
magnetic or electric charges for $m=(1,2,3 \dots )$.
The potential, between moduli and second rank gauge fields in (\ref{db6}), becomes
\be
V_2(\phi)=\ - \left [\ {\hat Q}_{\rm eff}^2 + Q^{(i)}D_{ij}Q^{(j)}\ \right ]\ ,\label{db150}
\ee
where ${\hat Q}_{\rm eff}$ and $Q^{(i)}$
denote the electric (or magnetic) charges, respectively, on the brane and in the effective string theory.

\par
\sp
Now, we consider a gauge choice $G_{i\a}=0$, for $(\a,\b)\equiv (x^4,x^1)$ and $(i,j)\equiv(x^2,x^3)$. Then,
the action (\ref{db6}) is simplified using a noncommutative scaling \cite{km-1}. The scaling incorporates 
vacuum field configurations for some of the fields and they are
\be
\partial_{\alpha}h_{ij}= 0\ , \quad R_{\bar h} = 0\ ,\quad \pr_{\a}\varphi^{(m)}=0\quad  {\rm and}\quad F^{(p)}_{\a\b}=0\ .\label{db151}
\ee
The relevant curved brane dynamics is governed by its on-shell action and is given by
\bea
&&S = -\int d^2x^{(\a)} d^2x^{(i)}
{\sqrt{\bar h}} {\sqrt{h}}\bigg [ {1\over{16\pi}}\left ( R_h - 2\Lambda\right )+
{1\over{64\pi}}h^{ij}\partial_i {\bar h}_{\alpha\beta}\partial_j{\bar h}_{\gamma\delta}
\epsilon^{\alpha\gamma}\epsilon^{\beta\delta}
\qquad\qquad\qquad\qquad {} \nonumber\\
&&\qquad\qquad\qquad\qquad\qquad\qquad\qquad\;\;\;\;\
-2 h^{ij}C_{mn}\pr_i\varphi^{(m)} \pr_j\varphi^{(n)}
- {1\over2} {\bar h}^{\alpha\beta} h^{ij} D_{pq}{\hat F}^{(p)}_{\alpha i}\star{\hat F}^{(q)}_{\beta j}
\bigg ]\ ,\label{db142}
\eea
where $C_{mn}$ and $D_{pq}$ for $p,q=(1,2,\ \dots i, i+1)$ are the appropriate moduli couplings and $\varphi^{(m)}$ take into account the dilaton and axions in the theory.

\section{Static dS and AdS geometries in $4D$}

\subsection{ Axially symmetric black holes}

In the case, we consider a constant scalar moduli in the theory (\ref{db142}) 
and restrict the EM-field on the brane only, 
$i.e.\ {\hat Q}\neq 0$ and $Q^{(i)}=0$. The anstaz for the gauge field 
${\hat A}_{\mu}\equiv ({\hat A}_t, {\hat A}_r, 0, 0)$ is given by
\be
{\hat A}_t = - {\hat Q}_{\rm eff} \sin\theta \cos\phi\; {\rm and}\qquad 
{\hat A}_r= {\hat Q}_{\rm eff} \sin\theta\sin\phi\ .\label{db105}
\ee
The non-vanishing components of the self-dual EM-field are
\be
{\hat E}_{\theta}= {\hat B}_{\theta} = {{{\hat Q}_{\rm eff}}\over{r}} \cos \phi\; {\rm and}\qquad 
{\hat E}_{\phi}= {\hat B}_{\phi}= - {{{\hat Q}_{\rm eff}}\over{r}}\sin \phi\ .\label{db106}
\ee
The general ansatz for a static and axially symmetric metric in the frame-work can be given by
\be
ds^2 =\ -f \ dt^2\ +\ f^{-1} h_1^2\ dr^2\ +\ f_0^{-1} h_1^2\ d\theta^2\
+\ f_0^{-1} h_2^2\ \sin^2\theta\ d\phi^2 \ ,\label{db11}
\ee
where ($f, f_0, h_1, h_2$) are arbitrary functions of ($r,\theta$).
Then, the independent components of Rici tensor in the theory can be expressed in terms of these arbitrary
functions and a constant potential $V_2(\varphi_0)$. The metric components are worked 
out to ${\cal O}(G^2_N)$ and they are given by
\bea
&&f=\ \left (k - {{\Lambda}\over{3}}r^2 - {{2\Lambda M_{\rm eff}}\over{r}}P_2(\cos \theta) + 
{{2C_{\rm eff}}\over{r^3}} - {{{\tilde C}_1}\over{r^3}}\sin^2 \theta\right )\ ,\nonumber\\
&&f_0= \ \left ( 1 + {{2C_{\rm eff}}\over{r^3}}  - {{{\tilde C}_2}\over{r^3}} \sin^2 \theta\right )\ ,\nonumber\\
&&h_1=\ \left (r^2 - {{2 C_{\rm eff}}\over{r}} + {{{\tilde C}_3}\over{r}} \sin^2 \theta \right )^{1/2}\nonumber\\
{\rm and}\qquad &&h_2=\ \left (r^2 - {{2 C_{\rm eff}}\over{r}} + {{{\tilde C}_4}\over{r}}\sin^2 \theta \right )^{1/2}\ ,\label{db111}
\eea
where $k$ takes constant values $(+1,0,-1)$ and governs the constant curvature geometry at the event horizon of
a black hole (\ref{db11}). The remaining parameters $C_{\rm eff}= (M_{\rm eff}Q^2_{\rm eff})$ and 
${\tilde C}_l$, for $l=(1,2,3,4)$, may be identified with the appropriate mass 
terms, of the order ${\cal O}(G^2_N)$, arising out of the EM-field in the theory.
Explicitly, the effective parameters governed by the ADM mass and charge of the black hole can be obtained from 
ref.\cite{km-2} and are given by
\bea
&&M_{\rm eff}=(G_NM)\left [ 1 - {{\Theta}\over{r^2}} + {\cal O}(\Theta^2)\right ]\qquad\qquad {}\nonumber\\
{\rm and}&&{\hat Q}_{\rm eff}= ({\hat Q}\sqrt{G_N})\left [ 1 - {{\Theta}\over{2r^2}} + {\cal O}(\Theta^2)\right ]
\ .\label{db131}
\eea
The black hole geometries (\ref{db11}) are characterized by four independent parameters 
($k$, $\Lambda$, $M_{\rm eff}$ and ${\hat Q}_{\rm eff}$) and they incorporate all order corrections in $\Theta$.
However, the black hole geometries appear to be sensitive to the order of Newton's constant in the theory. 
To ${\cal O}(G^2_N)$, the geometry governs an axially symmetric charged black hole, while to ${\cal O}(G_N)$,
it retains the spherical symmetry.
The horizon radius is determined by $f(r,\theta)=0$. Interestingly, there are three horizons 
in $dS_2\times S^2$ geometry. They are characterized by a cosmological horizon $r_c$, 
an event horizon $r_+$ and an inner horizon $r_-$. with ($r_c>r_+>r_-$). For ($M=0={\hat Q}$) and $\Lambda\neq 0$, 
the generic black hole geometry naively appears to 
describe a pure dS with a cosmological horizon at $r_c= (bk^{1/2})$, which is also formally  known as the dS radius. 
Similarly, for ($M\neq 0, {\hat Q}\neq 0$) and $\Lambda=0$, the geometry corresponds to a
generalized dS RN-like black hole with horizons at 
\be
r^{\rm dS}_{\pm} = M_{\rm eff} \pm \sqrt{M^2_{\rm eff} -k{\hat Q}^2_{\rm eff}}\ .\label{db0132}
\ee 
On the other hand, for $AdS$, the time coordinate turns out to be periodic $i.e.\ 
t_E\rightarrow t_E+2\pi R$, where $R=(bk^{1/2})$ denotes the AdS radius. For ($M=0={\hat Q}$) and $\Lambda\neq 0$, 
the generic black hole geometry corresponds to a pure AdS. For $\Lambda=0$, the event horizons are at 
\be
r^{\rm AdS}_{\pm} = M_{\rm eff} \pm \sqrt{M^2_{\rm eff} + k{\hat Q}^2_{\rm eff}}\ .\label{db0133}
\ee
The radius of the event horizon, for $k=1$, though resembles to that of a typical Schwarzschild black hole with
$r_h^{\rm AdS} \simeq 2M_{\rm eff}$, it governs a regular geometry there.

\sp

Now let us consider some of the interesting limits of an axially symmetric ansatz (\ref{db11}) which
leads to different kinds of black hole geometries in the frame-work. Naively, to order ${\cal O}(G_N)$,
the geometries describe the asymptotic dS and AdS black holes and retains the spherical symmetry.
They are
\bea
&&ds^2_{\rm dS} =\ -\left (k - {{r^2}\over{b^2}}\right) dt^2 +\  \left (k - {{r^2}\over{b^2}}\right)^{-1} dr^2 + r^2\ d\Omega^2\qquad\qquad {}\nonumber\\
{\rm and}
&&ds^2_{\rm AdS} =\ -\left (k + {{r^2}\over{b^2}}\right) dt^2 +\  \left (k + {{r^2}\over{b^2}}\right)^{-1} dr^2 + r^2\ d\Omega^2 \ .\label{db14}
\eea
Since the noncommutativity in the frame-work can be seen to redefine the polar angle $0\le \theta \le (\pi/n)$, 
for a large number  $n$ \cite{km-2}, the $\theta$ dependence drops out from the metric (\ref{db11}). 
It may alternately be achieved by  ${\tilde C}_l=0$  for an arbitrary polar angle $\theta$. A priori, the 
axially symmetric black hole geometries may be seen to retain its spherical symmetry in the frame-work. 
In fact, a careful analysis reveals that the black holes are described by their extremal geometries, 
$i.e.\ M\rightarrow 0$. In other words,
the black holes are governed by a spherical symmetry to ${\cal O}(G_N)$ only. The rotational symmetry comes into
play in the next order ${\cal O}(G^2_N)$. The correct extremal geometries, to ${\cal O}(G^2_N)$, are given by
\bea
&&ds^2_{\rm ds}=\ -l^2_{\perp}\left (k - {{r^2}\over{b^2}} + {{{\hat Q}^2_{\rm eff}}\over{r^2}}\right ) dt^2\ +\
\left (k - {{r^2}\over{b^2}} + {{{\hat Q}^2_{\rm eff}}\over{r^2}}\right )^{-1} dr^2\ +\
l^2_L \ d\Omega^2\nonumber\\
{\rm and} &&ds^2_{\rm Ads}=\ l^2_{\perp}
\left (k + {{r^2}\over{b^2}} - {{{\hat Q}^2_{\rm eff}}\over{r^2}}\right ) dt_E^2\ +\
\left (k + {{r^2}\over{b^2}} - {{{\hat Q}^2_{\rm eff}}\over{r^2}}\right )^{-1} dr^2\ +\
l^2_L \ d\Omega^2\ ,\label{db15}
\eea
where $l_{\perp}$ and $l_L=r_h$ are the associated length scales, respectively, in the transverse ($\perp$)- 
and in the longitudinal ($L$)-spaces.
It implies that the spherical symmetry in the metric is restored in the extremal limit.
Interestingly, for $k=1$, these solutions precisely correspond to the extremal black holes obtained from a
spherically symmetric ansatz considered in ref.\cite{kar06}.

\subsection{ Spherically symmetric black holes}

The generic black hole geometries may be obtained by retaining nontrivial moduli in the frame-work
This in turn can be seen to yield a nontrivial potential in the moduli space. Since the multiple
$U(1)$ gauge fields are coupled to the moduli, their presence can be seen to be significant in the
theory. Thus, in addition to the nonlinear $U(1)$ gauge field on the brane, there are multiple $U(1)$
gauge fields in the theory.
The anstaz for the non-vanishing (multiple) gauge fields $A^{(i)}$ are given by its $\perp$-components, 
and they are
\be
A^{(i)}_t= {{Q^{(i)}}\over{r}}\qquad {\rm and}\qquad A^{(i)}_{\phi}= Q^{(i)} \cos\theta\ .\label{db107}
\ee
The EM-field(s) are given by
\be
F^{(i)} = Q^{(i)}\left [{1\over{r^2}}\ dt\wedge dr + \sin\theta \ d\theta\wedge d\phi\right ]\ .\label{db108}
\ee
The non-vanishing electric or magnetic field components are given by $E_r^{(i)}=B_r^{(i)} 
= Q^{(i)}/r^2$. However the $E^2$ there, receives correction due to the multiple gauge fields at 
${\cal O}(G^2_N)$. It becomes
\be
{\hat E}^2\rightarrow E^2=\ {1\over{r^2}}\left ( {\hat Q}^2_{\rm eff}\ +\ {{G^2_N}\over{r^2}} Q^{(i)}D_{ij}Q^{(j)}\right )\ .
\label{db16}
\ee
In the case, the ansatz for a static, spherically symmetric, metric is given by
\be
ds^2 =\ - f \ dt^2\ +\ f^{-1}  dr^2\ + h^2 d\Omega^2\ ,\label{db109}
\ee
where $f$ and $h$ are arbitrary functions of $r$. The independent components of Ricci tensor in the theory
are worked out in an orthonormal basis to yield
\bea
&&R_{tt}=\ -{{f}\over{h^4}}V_{\rm eff}(\varphi)\ ,\nonumber\\
&&R_{rr}= \ 2 ( \partial_r\varphi)^2\ + {1\over{fh^4}} V_{\rm eff}(\varphi)\nonumber\\
{\rm and}&&R_{\theta\theta}=\ -{1\over{h^2}}V_{\rm eff}(\varphi)\
,\label{db110}
\eea
where $V_{\rm eff}(\varphi)$ denotes the effective potential in the moduli space due to all the gauge fields and
is a redefinition of $V_2(\phi)$ in eq.(\ref{db150}). It is given by 
$V_{\rm eff}(\varphi)=- {\hat Q}_{\rm eff}^{(p)}{D}_{pq}{\hat Q}^{(q)}_{\rm eff}$.
The arbitrary $f(r)$ is worked out to yield
\bea
f &=&\left ( k  -{{2 M_{\rm eff}}\over{r}} - {{\Lambda}\over3}r^2 \right)\left ( 1 \pm E^2\right )\ 
\nonumber\\
&=&\left ( k  -{{2 M_{\Theta}}\over{r}} - {{\Lambda}\over3}r^2 \right)\ ,\label{db132}
\eea
where
\bea
&&M_{\Theta} = M_{\rm eff}\left (1\pm {{{\hat Q}^2_{\rm eff}}\over{r^2}} \pm {{G^2_N}\over{r^4}} Q^{(i)} D_{ij}
Q^{(j)} + \ \dots\ \right )\qquad\qquad\qquad {}\nonumber\\ 
&&\qquad\qquad\qquad\qquad\qquad\mp \left (k - {{\Lambda}\over3}r^2\right ) 
\left ( {{{\hat Q}^2_{\rm eff}}\over{2r}} + {{G^2_N}\over{2r^3}} Q^{(i)}D_{ij}Q^{(j)} +\ \dots\ \right )
\ .\label{db133}
\eea
The moduli significantly modify the radius of $S^2$ due to its coupling to the EM-field in the theory.
The electric (magnetic)  charges there reduce the radius of otherwise $S^2$ and the effective radius can 
be expressed as
\be
h =\ r \ e^{\varphi(r)}\ ,\quad {\rm where}\quad e^{\varphi(r)}= \left ( e^{2\varphi_h} -
{{G_N}\over{r r_h}} Q^{(i)}D_{ij}Q^{(j)}\right )^{1/2}\ ,\label{db18}
\ee
where the constant $\varphi_h$ is the value of $\varphi$ at the event horizon $r_h$.
Then, the generic dS RN-like black hole geometry, to ${\cal O}(G_N)$ in the theory, is given by
\bea
&&ds^2 = - \left ( k - {{r^2}\over{b^2}}- {{2M_{\rm eff}}\over{r}} + {{G_N{\hat Q}^2}\over{r^2}} - 
{{\Theta {\hat Q}_{\rm eff}^2}\over{r^4}}\right ) dt^2\qquad\qquad\qquad {}\nonumber\\
&&\qquad\qquad\qquad\qquad
+ \left ( k- {{r^2}\over{b^2}} - {{2M_{\rm eff}}\over{r}} + {{{G_N}{\hat Q}^2}\over{r^2}} - 
{{\Theta {\hat Q}_{\rm eff}^2}\over{r^4}}\right )^{-1} dr^2\ +\  e^{2\varphi(r)} \ r^2 d\Omega^2\ .\label{db19}
\eea
It implies that the area of the event horizon tend to shrink in presence of moduli in the theory and the
reduction in the horizon radius begins even at ${\cal O}(G_N)$. On the other hand, the moduli corrections 
along ($t,r$)-space to the  black hole geometry are at ${\cal O}(G^2_N)$. It is straight-forward to check
that the $dS$ geometry (\ref{db19}) reduces to the one obtained in an effective string theory \cite{garfinkle}
in absence of a $D_3$-brane and with ($k=1, \Lambda=0$). 

\sp
\noindent
Like-wise, the generic AdS RN-like black hole geometry, to ${\cal O}(G_N)$, is worked out to yield
\bea
&&ds^2 = \left ( k + {{r^2}\over{b^2}}- {{2M_{\rm eff}}\over{r}} - {{G_N{\hat Q}^2}\over{r^2}} + {{\Theta {\hat Q}_{\rm eff}^2}\over{r^4}}\right ) dt_E^2\qquad\qquad\qquad {}\nonumber\\
&&\qquad\qquad\qquad\qquad
+ \left ( k + {{r^2}\over{b^2}} - {{2M_{\rm eff}}\over{r}} - {{{G_N}{\hat Q}^2}\over{r^2}} + 
{{\Theta {\hat Q}_{\rm eff}^2}\over{r^4}}\right )^{-1} dr^2\ +\  e^{2\varphi(r)} \ r^2 d\Omega^2\ .\label{db20}
\eea
In presence of $\Theta$ corrections, the interesting feature of the
gravity decoupling limit $g\rightarrow 0$ can be revisited in the formalism. In the limit, the theory is
described by its low energy string modes and hence, essentially governs a semi-classical regime. In fact, the
limit can equivalently be obtained by taking $M\rightarrow 0$. A priori, the extremal 
$dS_2\times S^2$ and $AdS_2\times S^2$ geometries can be seen to be governed by $4D$ black holes.
However the correct geometry, in the gravity decoupling limit, turns out to be that of
an emerging two dimensional semi-classical theory within a curved $D_3$-brane \cite{km-1,km-2,km-3,kar06}.
In other words, the $\Theta$ term acts as a propagator in an internal space between the
quantum and classical modes in the theory. It scales away the relevant quantum modes along the $S^2$ and
leaves behind the semi-classical gravity along the $\perp$-space of the brane-world.{\footnote{Originally, the
scaling analysis in $4D$ Einstein's gravity, at Planck scale, was developed by Verlinde and Verlinde 
\cite{ver-2}. Interestingly, in a collaboration with Maharana \cite{kar-maharana}, the scaling analysis
was explored to study the scattering of non-abelian gauge particles at Planck scale.}}Then the extremal 
dS and AdS black hole geometries are precisely described by the eqs.(\ref{db15}). In the context, the
Hawking temperature, $T_H^{(k)}= {1\over{4\pi}} [\pr_rf(r)]_{r\rightarrow r_h}$, is computed for a 
$4D$ black hole (\ref{db20}) and is given by
\be
T_H^{(k)}= {{3r_h^2 + kb^2}\over{4\pi b^2 r_h}}\ .\label{db205}
\ee
It implies that that Hawking temperature for different horizon geometries, associated with $k=(+1,0,-1)$, satisfy
$T_H^{(1)}>T_H^{(0)}>T_H^{(-1)}$. In other words, the Hawking temperature for an AdS Schwarzschild black hole
is bounded from below.

\section{Extremal geometries and extra dimensions}

\subsection{Two dimensional black holes}

The fact that $S^2$ shrinks to zero size, $i.e.\ h\rightarrow 0$, in the semi-classical regime can 
be incorporated naively into the frame-work to yield
\be
\varphi_h = {1\over2} \ln \left ( {G_N} Q^{(i)}D_{ij}Q^{(j)}\right ) - \ln r_h\ ,\label{db22}
\ee
Then, the emerging $dS_2$ and $AdS_2$ geometries is obtained in the extremal limit from eqs.(\ref{db19}) 
and (\ref{db20}). They are, respectively, given by
\bea
&&ds^2= - \left (k - {{r^2}\over{b^2}}  + {{{\hat Q}_{\rm eff}^2}\over{r^2}} \right) dt^2 +
\left (k - {{r^2}\over{b^2}} + {{{\hat Q}_{\rm eff}^2}\over{r^2}} \right)^{-1} dr^2 \nonumber\\
{\rm and}&&ds^2= \left (k + {{r^2}\over{b^2}}  - {{{\hat Q}_{\rm eff}^2}\over{r^2}} \right) dt_E^2 +
\left (k + {{r^2}\over{b^2}} - {{{\hat Q}_{\rm eff}^2}\over{r^2}} \right)^{-1} dr^2 \ .\label{db21}
\eea
The event horizon equation for an dS geometry is worked out, in presence of back reaction, for its horizon radii. 
They can be seen to satisfy
\bea
&&h^2= b {\hat Q}_{\rm eff}\ ,\qquad {\rm for}\quad k=0\ .\nonumber\\
{\rm and} \quad
&&h_-^2 = {{{\hat Q}^2_{\rm eff}}\over{k}}\ ,
\quad h_+^2= \left (b^2k + {{{\hat Q}^2_{\rm eff}}\over{k}}\right )\quad {\rm for}\quad k\neq 0
\ .\label{db23}
\eea
Then, the relation between the moduli and charges (\ref{db22}) is modified due to the nonlinear EM-field
contribution. Using eqs.(\ref{db18}) and (\ref{db23}), one finds
\bea
&&\varphi_h= {1\over2} \ln \left ( b{\hat Q}_{\rm eff} + {G_N} Q^{(i)}D_{ij}Q^{(j)}\right ) -\ \ln r_h\qquad {\rm for}\;\ k=0 \qquad {}\nonumber\\
{\rm and}&&\varphi_{h_-} = {1\over2} \ln \left ( {{\hat Q}^2_{\rm eff}}\ + \ (k{G_N} )\ 
Q^{(i)}D_{ij}Q^{(j)}\right ) -\ \ln (kr_h)\ ,\nonumber\\
&&\varphi_{h_+} = {1\over2} \ln \left ( b^2k + {{\hat Q}^2_{\rm eff}}\ +\ (k{G_N})\ 
Q^{(i)}D_{ij}Q^{(j)}\right ) - \ \ln (kr_h)\qquad {\rm for}\;\ k\neq 0\ .\label{db24}
\eea
A similar analysis is performed for the AdS black hole and the moduli can be shown to be expressed in terms of
the EM-charges in the frame-work.
It implies that the value of moduli at the event horizon can be expressed in terms of the 
electric (magnetic) charges there. It reassures the fact that the event horizons of dS and AdS black holes are attractors in a curved brane-world frame-work as well.
It is interesting to note that the $2D$ extremal black holes (\ref{db21}), for $k\neq 0$, interchange between dS and AdS geometries under $k\leftrightarrow -k$. This is accompanied by the fact that their respective 
topologies get interchanged, $i.e.\ \Re \times S^1 \leftrightarrow S^1\times \Re$, which may be seen to be 
independent of the value of $k$. For instance, the semi-classical dS and AdS geometries can be well approximated by large $r$ to
yield
\be
ds^2 = \left (- r^2 dt^2 +\ {{dr^2}\over{r^2}}\right )\ .\label{db25}
\ee
The geometry is indeed independent of the value of $k$. This is possibly of no surprise, as $k$ determines the
geometry associated with a constant curvature event horizon.

\subsection{Hawking radiation and geometric transitions}

\FIGURE[pos]{\epsfig{file=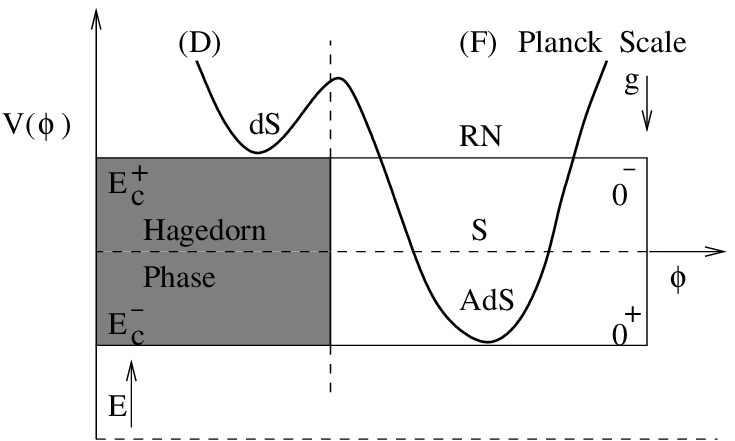}
\caption{The variation of $V(\phi)$ in the moduli space leading to dS and AdS vacua at its
minima, respectively, in the EM-string and in the fundamental bosonic string sectors.}}


Let us begin this section by briefly revisiting the Hawking radiation phenomenon in 
$dS_4$ (\ref{db19}) and $AdS_4$ (\ref{db20}) black holes leading to some of the plausible 
extremal geometries in the frame-work. Interestingly, the nonlinear EM-field in a curved $D_3$-brane 
frame-work may facilitates one to view the radiation as a result of some pair production processes 
in vacuum. For instance, a member from a pair created just outside an event horizon tunnels to inside 
at the expanse of zero total energy. The member moving inside and away from the event horizon can 
equivalently be viewed as a member, with an opposite charge, moving out of the horizon.
Though a member in a primary pair is time dilated with respect to the other pair outside the horizon, 
it can undergo a pair annihilation with an opposite charge member of the secondary pair to produce a 
quanta of Hawking radiation. The pair production process continues with an increase in $E$-field until 
the critical value $E_c$ is reached. The radiation phenomenon is primarily based on the global conservation of energy-momentum and, a priori, it appears to be independent of the thermal analysis. However, our recent
analysis \cite{kar06} suggests that thermal mechanisms do play a role to decouple the condensate of gauge nonlinearity from the AdS vacua in string theory. In other words, the decoupling of tachyon condensate \cite{kar-panda}
in an open bosonic string theory, leads to the dS vacua on the brane-world.


\par
\sp

In the gravity decoupling regime, $i.e.$ in the matter dominated brane-world, the figure 1 depicts the fact that 
the effective space-time is governed by the 
dS vacua in the Hagedorn phase of EM-string. Interestingly, the dS vacua can tunnel to AdS vacua in the string theory.
Though, the Hagedorn phase is at the sub-Planckian regime from a reference zero-point energy defined in the underlying 
EM-string theory, within the $D$-brane-world, it depicts the low energy regime of the fundamental string theory. 
For instance, the minima in AdS vacua may correspond to a five dimensional 
Schwarzschild like black hole for $k=1$, which Hawking
radiates to yield a typical $4D$ Schwarzschild black holes at $S$. Hawking radiation further incorporates
geometric transitions to yield RN-black holes, which is presumably described by the dS vacua.\label{ds-adsf}.

\sp
Now, incorporating the decoupling of the gravity as well as the gauge nonlinearity,
the black hole geometries (\ref{db19}) and (\ref{db20}) reduce to their near
horizon geometries. The naive $dS_2$ geometry describes a typical monopole black hole solution and is
given by
\be
dS^2 = - \left ( k - {{r^2}\over{b^2}} - {{\mu}\over{r^2}} \right ) dt^2 +
\left ( k - {{r^2}\over{b^2}} - {{\mu}\over{r^2}} \right )^{-1} dr^2\ ,\label{db26}
\ee
where $\mu= -(G_N{\hat Q}^2)$ is the reduced mass of the black hole. The gravitational potential
generated by $\mu$ leads to a central force, which in turn is governed by $1/r^3$ law and hints at the existence of
an underlying theory of gravity in five dimensions. It favors the assertion of three extra large 
dimensions in addition to the $2D$ black hole geometry within a curved $D_3$-brane frame-work
\cite{km-2,km-3}. Incorporating the required large extra dimensions one may alternately elevate 
$2D$ near horizon geometry to an appropriate $dS_2\times S^3$ or an $AdS_2\times S^3$. The idea 
is in agreement with the $dS_5$/CFT construction \cite{strominger2} in quantum gravity and 
with the established $AdS_5$/CFT correspondence in string theory \cite{maldacena}. 

\par
\sp
On the other hand, the observed tunneling between dS and AdS vacua can be explained by analyzing the behavior
of Hawking temperature (\ref{db205}), for $k=1$, $T_H= (1/4\pi) [ r_h^{-1} - \Lambda r_h ]$ in presence of 
radiation.
For instance an $AdS_4$ RN-like black hole has been shown to possess to a Schwarzschild radius and 
can be obtained from the vacuum solution to the $5D$ effective theory  In the regime, 
the $T_H$ increases with the Hawking radiation and reaches a local maxima, where the AdS geometry transforms 
to the dS RN-like black hole. Subsequently, $T_H$ drops gradually to zero with the radiation and is governed 
by an extremal RN-like geometry. Further decoupling of gauge nonlinearity is favored leading to a rise in 
$T_H$ to the Hagedorn temperature. There, the Hawking radiation ceases and 
the extremal RN black hole was identified with a monopole solution. On the other hand, for large $r$, the
black hole solution reduces to a pure dS geometry.
The analysis suggests that a presumably stable AdS vacuum in a curved brane theory exhibits quantum tunneling 
and can make its way to dS vacuum in the low energy regime.

\subsection{Five dimensional perspective in curved brane-world}

In principle the $AdS_5$ gravity, in the string bulk, may be obtained from a higher dimensional 
theory, such as $M$-theory \cite{banks,ikkt} or $F$-theory \cite{vafa}. Since a $D_3$-brane is
described in a type IIB string theory, we consider a compactification on $K^3\times S^1$ there.
The $5D$ effective string action in Einstein frame may be given by
\be
S= - \int d^5 x \sqrt{{\tilde G}^E}\ \left [{1\over{16\pi {\tilde G}_N}}
\left ({\tilde R} - 2\tilde\Lambda \right ) -\ {1\over3}(\partial\Phi)^2 -
\ {1\over2} {\tilde F}_1^{(k)}{\cal C}_{kl}{\tilde F}_1^{(l)}
- \ {1\over4}{\tilde F}_2^{(p)}{\cal D}_{pq} {\tilde F}_2^{(q)}\right ]
\ .\label{db55}
\ee
where ${\tilde G}_N$ denotes the Newton constant, $\Phi$ is the dilaton and ${\tilde Z}$ is
a normalization constant in $5D$.
The moduli matrices ${\cal C}_{kl}$ and ${\tilde{\cal D}}_{pq}$
describe their couplings to the gauge fields in the theory.
The effective cosmological term in the action
is generated by the multiple five-form field strengths. It is
given by
\be
\tilde\Lambda =\ {{\pi{\tilde Z}{\tilde G}_N}\over{30}} {\tilde F}_5^{(m)}{\cal L}_{mn}{\tilde F}_5^{(n)}\ .
\label{db56}
\ee
At the extrema of the potential , the moduli ${\cal L}_{mn}$ take a constant value.
Thus for stable moduli, ${\tilde\Lambda}$ can be interpreted as a cosmological constant in $5D$ effective
string theory (\ref{db55}). The multiple zero-forms in the $5D$ theory 
can make $\tilde\Lambda= -6/b^2$ large. This in turn would give rise to a small periodicity $\beta= 2\pi b$ 
in the time coordinate unlike to that in $4D$.
Interestingly, for ${\tilde\Lambda}=0$, the black hole solutions in the curved brane frame-work were 
obtained in a collaboration \cite{km-3}. They are shown to describe the gravity dual proposed for $D_3$-brane 
in presence of a constant magnetic field \cite{li-wu}. However for $\tilde\Lambda\neq 0$, one can check that
an appropriately elevated geometry (\ref{db26}) to $AdS_5$, satisfies the equations of motion in the bulk theory (\ref{db55}). In fact, the $AdS_5$ geometry can be formally re-expressed in terms of $dS_5$ by $k\rightarrow -k$
and when necessary, it can appropriately be transformed back to $AdS_5$. 
Interestingly, the $dS_5$ monopole black hole in the 
frame-work is characterized by its ADM mass ${\tilde M}= -({\hat Q}^2/4)$ and may be given by
\be
ds^2= - \left (k- {{r^2}\over{b^2}}  - {{4G_N{\tilde M}}\over{r^2}}\right) dt^2 +
\left ( k- {{r^2}\over{b^2}} + {{4G_N{\tilde M}}\over{r^2}} \right)^{-1} dr^2 + r^2\ d\Omega_3^2\ .\label{db27}
\ee
The generic line element on $S^3$ may be given by
\be
d\Omega_3^2 = \ d\rho^2 +\ R_k^2 \ d\Omega^2\ ,\label{db28}
\ee
where $R_k$ is the radius of $S^2$ and can be given by
\be
R_0=\ \rho\ ,\qquad R_1=\ \sin \rho\quad {\rm and} \qquad R_{-1} =\ \sinh \rho\ .\label{db29}
\ee
It is straight-forward to check that the event horizon (\ref{db27}) is described, respectively, by a flat, an 
elliptic and an hyperbolic geometry, for $k=0$, $k=1$ and $k=-1$.
However, the horizon radii (\ref{db23}) for the $dS_5$ black hole are worked out to yield event horizon only
for $k=1$ and $k=-1$, which are, respectively, governed by $r_+= (b^2 -4{\tilde M})^{1/2}$ and 
$r_-=2{\tilde M}^{1/2}$ for $k=-1$. The other horizons, including $k=0$ case, turn out to be unphysical there.
Analysis reveals that the $dS_5$ black hole event horizon is at $r_+$ and the $AdS_5$ possesses a regular geometry
with an event horizon at $r_-$. The Hawking temperature, in the case, turns out to become
\be
{\tilde T}_H^{(k)} = {{2r_h^2 + kb^2}\over{2\pi b^2r_h}}\ .\label{db30}
\ee
Thus, the expression for Hawking temperature is modified in presence of ${\tilde\Lambda}$. For small black holes,
the Hawking temperature can be approximated by that in absence of cosmological constant. However for large black holes,
the contribution due to the cosmological constant becomes significant role to define $T_{H}$ in the theory.

\par
\sp

Since the semi-classical gravity is obtained in the low energy limit
of a string theory, the $5D$ monopole black hole geometry (\ref{db27}) essentially governs a pure dS vacuum. Interestingly, when the same geometry is viewed in its gravity dual, $i.e.$ a strongly coupled gauge, theory, 
it corresponds precisely to a monopole, with an $U(1)$ magnetic charge ${\hat Q}$, in an asymptotically flat 
geometry. It suggests that the gauge nonlinearity is completely accountable to the value of
cosmological constant in the $4D$ brane-world.

\subsection{Higher dimensional geometries}

The extremal $dS_4$ and $AdS_4$ geometries leading to $2D$ black holes (\ref{db21}) can be seen to be a 
potential candidate to shed light on the extra dimensions in a curved brane-world. For simplicity, we begin 
with an $AdS_2$ Schwarzschild black hole geometry. In particular, we restrict the back reaction of the metric,
predominantly, to be first order in $\Theta$ correction to its reduced mass. Then, the time component of the 
metric, just before $\Theta$ decoupling may be given by
\be
f_1= \left ( 1 + {{r^2}\over{b^2}} - {{\mu_1}\over{r^4}}\right )\ .\label{db31}
\ee
where the reduced mass $\mu_1= -\Theta{\hat Q}^2_{\rm eff}$. 
The gravitational potential in the case leads to
a central force and can be seen to be governed by $1/r^5$ law, which in turn implies a $7D$ space-time there. 
Like wise, for higher orders
in $\Theta$, it may be possible to argue for the existence of arbitrary (odd) dimensional manifold, such as
$D=(9,11,\ \dots )$ in the extremal limit of a large black hole. The reduced masses 
($\mu_2=\Theta^2{\hat Q}_{\rm eff}$, $\mu_3= \Theta^3 {\hat Q}_{\rm eff}, \dots $) 
re-generated in each step, are defined by a non-perturbative charge. They satisfy $\mu_1>\mu_2>\dots \ $, for a fixed $|\Theta|<1$. In other words, the potential
takes its minimum value, which in turn would set-up an upper bound to the dimensions of space-time geometries.
Similarly, some of the even dimensional $D=(6,8,10,12.\ \dots )$ geometries become prominent, when the analysis is
worked out for a small uncharged black hole. In that case, the reduced mass 
${\tilde\mu}_1=\Theta M_{\rm eff}$ is governed by a central force $1/r^4$ and shows the existence of a $6D$ and 
subsequently higher (even) dimensional geometries. It implies that the dimension of space-time may well be
governed by an effective potential $V(\Theta)$ in the moduli space and $D=(4,5 \dots 10,11)$ and $D=12$ correspond
to various stable extrema there.

\section{Summary}

To summarize, we have obtained generic de Sitter and Anti de Sitter static black hole geometries in $4D$ in a 
noncommutative brane-world. In particular, a constant moduli case was
investigated to yield an axially symmetric dS and AdS black hole geometries with different horizon topologies.
In addition, a case with an arbitrary moduli was analyzed for the spherically symmetric black holes in dS and AdS spaces.
These black holes were shown to be characterized by their generalized ADM mass $M_{\rm eff}$, generalized electric or
magnetic charge ${\hat Q}_{\rm eff}$, in addition to ($k,\Lambda$). The generalized mass and charge were shown to
incorporate all orders in noncommutative parameters $\Theta$. However, the ${\cal O}(G_N)$ was argued to be sensistive
to the symmetry of a black hole solution in the frame-work. For instance, it was shown that an axially symmetric black hole
becomes significant to ${\cal O}(G^2_N)$, while the spherically symmetry was restored to ${\cal O}(G_N)$.
It was argued that the gauge nonlinearity in the theory enforces back reaction into the metric. 
The multiple gauge fields in the frame-work were shown to incorporate  a shrink in the radius of event horizon. 
The extremal dS and AdS with different horizon geometries with 
a constant curvature were obtained. It was shown that the axially symmetric solution retains its spherical symmetry
in the extremal limit. The moduli in the theory were shown to be governed by the electric (magnetic) charges at the
event horizon of the dS and AdS black holes, which in turn reconfirms the attractor behavior of event horizon in the
frame-work. In addition, the emerging notion of two dimensional de Sitter monopole black hole in the curved brane 
frame-work is significant. Thus, the $dS_2$ vacuum appears to be a potential candidate to explore some of the 
quantum gravity aspects in the frame-work.

\par
\sp
It was shown that the event horizon of the extremal AdS coincides with a curvature singularity in the dS space,
leading to a single geometry in $2D$. The effective gravitational potential associated with the reduced mass of extremal dS black hole was analyzed to confirm the presence of three extra large dimensions in the regime. 
The Hagedron transitions in the near horizon geometry
of $dS_5$ monopole black hole was analyzed to decouple the $\Theta$-terms. The new $dS_5$ vacuum with a nontrivial
cosmological potential possibly describes our brane-world, $i.e.$ a $D_3$-brane, at its boundary. The potential
was argued to be at its local minima on the brane-world  and describes a small positive constant $\Lambda$. Different
space-time dimensions may be guided by different vacua at its extremum in the moduli space.

\par
\sp
It may be interesting to extend the curved brane-world, inspired by a noncommutative $D_5$-brane in type IIB string
theory or in a $D=12$ construction for low energy gravity theory. With an appropriate choice of nonlinear EM-field,
it may be possible to obtain $dS_4$ and $AdS_4$ black hole geometries using the noncommutative scaling in the theory.
Then, the emerging large $4D$ black hole may provide some concrete clue to the tunneling between $dS$ and $AdS$ vacua 
in the low energy string theory.

\acknowledgments

I would like to thank the Abdus Salam I.C.T.P, Trieste for hospitality, where this work
was completed. The research was supported in part by the D.S.T., Govt.of India, under SERC fast track 
young scientist project PSA-09/2002.

\def\anp{Ann. of Phys.}
\def\cmp{Commun. Math. Phys.}
\def\atmp{Adv.Theor.Math.Phys.}
\def\prl{Phys. Rev. Lett.}
\def\prd#1{{Phys. Rev.} {\bf D#1}}
\def\jhep{J.High Energy Phys.}
\def\cqg{{Class. \& Quantum Grav.}}
\def\plb#1{{Phys. Lett.} {\bf B#1}}
\def\npb#1{{Nucl. Phys.} {\bf B#1}}
\def\mpl#1{{Mod. Phys. Lett} {\bf A#1}}
\def\ijmpa#1{{Int. J. Mod. Phys.} {\bf A#1}}
\def\ijmpd#1{{Int. J. Mod. Phys.} {\bf D#1}}
\def\rmp#1{{Rev. Mod. Phys.} {\bf 68#1}}



\begin{thebibliography}{99}

\baselineskip= 18 truept

\bibitem{witten1}E. Witten, {\tt hep-th/0106109} (2001).

\bibitem{strominger2}A. Strominger, \jhep{\bf 10}, 029 (2001).

\bibitem{kklt}S. Kachru, R. Kallosh, A. Linde and S.P. Trivedi, \prd{\bf 68}, 046005 (2003).

\bibitem{maldacena}J. Maldacena, \atmp{\bf 2}, 231 (1998).

\bibitem{witten2}E. Witten, \atmp{\bf 2}, 253 (1998).

\bibitem{gibbons-ishi}G.W. Gibbons and A. Ishibashi, \cqg {\bf 21}, 2919 (2004).

\bibitem{jevicki}A. Jevicki and S. Ramgoolam, \jhep{\bf 04}, 032 (1999).

\bibitem{ardalan}F. Ardalan, H. Arfaei, M.R. Garousi and A. Ghodsi, \ijmpa {\bf 18}, 1051 (2003).

\bibitem{vassilevich}D.V. Vassilevich, \npb{\bf 715}, 695 (2005).

\bibitem{nasseri}F. Nasseri, Gen.Rel.Grav.{\bf 37}, 2223 (2005).

\bibitem{km-1}S. Kar and S. Majumdar, \ijmpa{\bf 21}, 2391 (2006).

\bibitem{nicolini}P. Nicolini, A. Smailagic and E. Spallucci, \plb{\bf 632}, 547 (2006).

\bibitem{km-2}S. Kar and S. Majumdar, {\tt hep-th/0510043} (In press, \ijmpa {}, 2006).

\bibitem{cho2}I. Cho, E.J. Chun, H.B. Kim and Y. Kim, {\tt hep-ph/0601147} (2006) .

\bibitem{horowitz2}G.T. Harowitz and J. Polchinski, {\tt gr-qc/0602037} (2006).

\bibitem{aschieri}P. Ascieri, M. Dimitrijevic, F. Meyer and J. Wess, \cqg {\bf 23}, 1883 (2006).

\bibitem{alvarez}L. Alvarez-Gaume, F. Meyer and M. Vacquez-Mozo, {\tt hep-th/0605113} (2006).

\bibitem{km-3}S. Kar and S. Majumdar, {\tt hep-th/0606026} (2006).

\bibitem{lopez}J.C. Lopez-Diminguez, O. Obregon, M. Sabido and C. Rmirez, {\tt hep-th/0607002} (2006).

\bibitem{kar06}S. Kar, {\tt hep-th/0607029} (2006).

\bibitem{sazbo}R.J. Szabo, {\tt hep-th/0606233} (2006) and the references there in.

\bibitem{seiberg-witten}N. Seiberg and E. Witten, \jhep{\bf 09}, 032 (1999).

\bibitem{li-wu}M. Li and Y.-S. Wu, \prl{\bf 84}, 2084 (2000).

\bibitem{hashi-itzhaki}A. Hashimoto and N. Itzhaki, \plb{\bf 465}, 142 (1999).

\bibitem{ho-li}P.-M. Ho and M. Li, \npb{\bf 596}, 259 (2001).

\bibitem{vafa}C. Vafa, \npb{\bf 469}, 403 (1996).

\bibitem{tseytlin}A. A. Tseytlin, \prl{78}, 1864 (1997).

\bibitem{kar97}S. Kar, \npb{\bf 497}, 110 (1997).

\bibitem{pope}N. Khviengia, Z. Khvienga, H. Lu and C.N. Pope, \cqg{\bf 15}, 759 (1998).

\bibitem{hewson}S.F. Hewson, \npb{\bf 534}, 513 (1998).

\bibitem{garfinkle}D. Garfinkle, G.T. Horowitz and A. Strominger, \prd{\bf 43}, 3140 (1991), Erratum, ibid.{\bf D45},
3888 (1992).

\bibitem{ver-2}E. Verlinde and H. Verlinde, \npb{\bf 371}, 246  (1992).

\bibitem{kar-maharana}S. Kar and J. Maharana, \ijmpa{\bf 10}, 2733 (1995).

\bibitem{kar-panda}S. Kar and S. Panda, \jhep{\bf 11}, 052 (2002).

\bibitem{banks}T. Banks, W. Fishler, S.H. Shenker and L. Susskind, \prd {\bf 55}, 5112 (1997)

\bibitem{ikkt}N. Ishibashi, H. Kawai, Y. Kitazawa and A. Tsuchiya, \npb{\bf 498}, 467 (1997).


\end{thebibliography}
\end{document}